\documentclass{article}

\usepackage[preprint, nonatbib]{neurips_2020}
\usepackage[utf8]{inputenc} 
\usepackage[T1]{fontenc}    
\usepackage{hyperref}       
\hypersetup{
    colorlinks = true
}
\usepackage{url}            
\usepackage{booktabs}       
\usepackage{amsfonts}       
\usepackage{nicefrac}       
\usepackage{microtype}      
\usepackage{cleveref}
\usepackage{graphicx}
\usepackage[ruled,vlined]{algorithm2e}
\title{Echo Chambers in Collaborative Filtering Based Recommendation Systems
}

\author{%
    Emil Noordeh$^{1,*}$, Roman Levin$^{2,*}$, Ruochen Jiang$^{3,*}$, Harris Shadmany$^{4,*}$\\
    $^1$Stanford University\\ $^2$University of Washington\\ $^3$University of California, Los Angeles\\ $^4$University of California, Berkeley\\
    $^*$All authors contributed equally to this work

}

\begin{document}

\maketitle

\begin{abstract}
Recommendation systems underpin the serving of nearly all online content in the modern age. From Youtube and Netflix recommendations, to Facebook feeds and Google searches, these systems are designed to filter content to the predicted preferences of users. Recently, these systems have faced growing criticism with respect to their impact on content diversity, social polarization, and the health of public discourse. In this work we simulate the recommendations given by collaborative filtering algorithms on users in the MovieLens data set. 
We find that prolonged exposure to system-generated recommendations substantially decreases content diversity, moving individual users into ``echo-chambers'' characterized by a narrow range of content. Furthermore, our work suggests that once these echo-chambers have been established, it is difficult for an individual user to break out by manipulating solely their own rating vector.

\end{abstract}

\section{Introduction}
\indent 
Recommendation systems can have drastic effects on user opinion formation and content preference on large scales. Consumers in recommendation systems have been found to experience ``preference feedback loops'', whereby comparisons to others who share their preferences results in recommendations of greater homogeneity \cite{bobadilla2013recommender}. This is one manifestation of recommendation systems having unintended side effects, namely the creation of “filter bubbles” and “echo chambers” \cite{dandekar2013biased}. 

Google DeepMind recently conducted a theoretical study of the impact of collaborative filtering on content diversity and found that ``degenerate feedback loops” are an intrinsic feature of recommender systems \cite{jiang2019degenerate}. These collaborative filtering techniques work by creating user recommendations based on the preferences of similar users in the system. The DeepMind study found that ``random exploration” by a ``large candidate pool” is effective in slowing and combating system degeneracy. This suggests that there is a need for further research and exploration of model design.

Recommender systems used in production at large scale almost all utilize collaborative filtering techniques in some way. While these techniques do a good job of accurately predicting the content users might enjoy, accuracy may not be the only metric on which to gauge the success of a model. McNee et al. in \cite{mcnee2006being} highlight the importance of user-centric metrics which gauge different aspects of the quality of recommendation lists such as content diversity.

\indent 
This work aims to answer two central questions:
\begin{enumerate}

\item {\bf Do recommender systems substantially reduce content diversity and contribute to echo chamber formation over time?} 
\item {\bf If so, is it possible for a user to break out of the echo chamber by manipulating solely their own content ratings?}
\end{enumerate}

 In the following sections we leverage the MovieLens dataset to simulate users in a recommendation system. We visualize and quantitatively measure the impact of these systems on the diversity of content presented to the users.
 
\section{Data and Methods}

\subsection{Movielens Dataset}\label{movie_lens}
The MovieLens dataset contains the ratings of $\sim28$ million users across $\sim58$ thousand movies \cite{harper2015movielens}. A feature of this dataset that is critical for our study is the concept of ``movie-tags" \cite{vig2012tag}. These tags encode each movie in an information space characterized by terms like ``futuristic", ``scenic", and ``humorous". They were created through community-supervised learning, a learning model trained on user data (namely tag exploration, text reviews, and ratings) \cite{vig2012tag}. The relevance of each movie-tag to each movie is provided in the dataset as a continuous value between 0 (irrelevant) and 1 (very relevant). There are a total of 1128 movie-tags that are used to characterize the content of a film and we leverage these to compute the diversity of a set of movies as discussed in \cref{sec:diversity}. Since we are interested in using the movie-tags to measure content diversity, we filter the Movielens data to only include movies that have movie-tag information. There are 13176 such movies.

\subsection{Simulating recommendation systems}\label{sec:sim}

We took an experimental approach to studying the evolution of the diversity of recommended content over time. We simulated a recommendation system and repeated user interactions with it and analyzed the effect of the prolonged exposure to the system on the diversity of the recommendations. In our experiments, we focus on collaborative filtering based recommender systems. Specifically, we used the SVD model described in \cite{koren2009matrix} and section 5.3.1 of \cite{Ricci_book} -- a matrix factorization method that models user-item interactions as inner products in a low-rank latent space. 

\begin{figure}[!htb]
\centering
\includegraphics[width=0.9\textwidth]{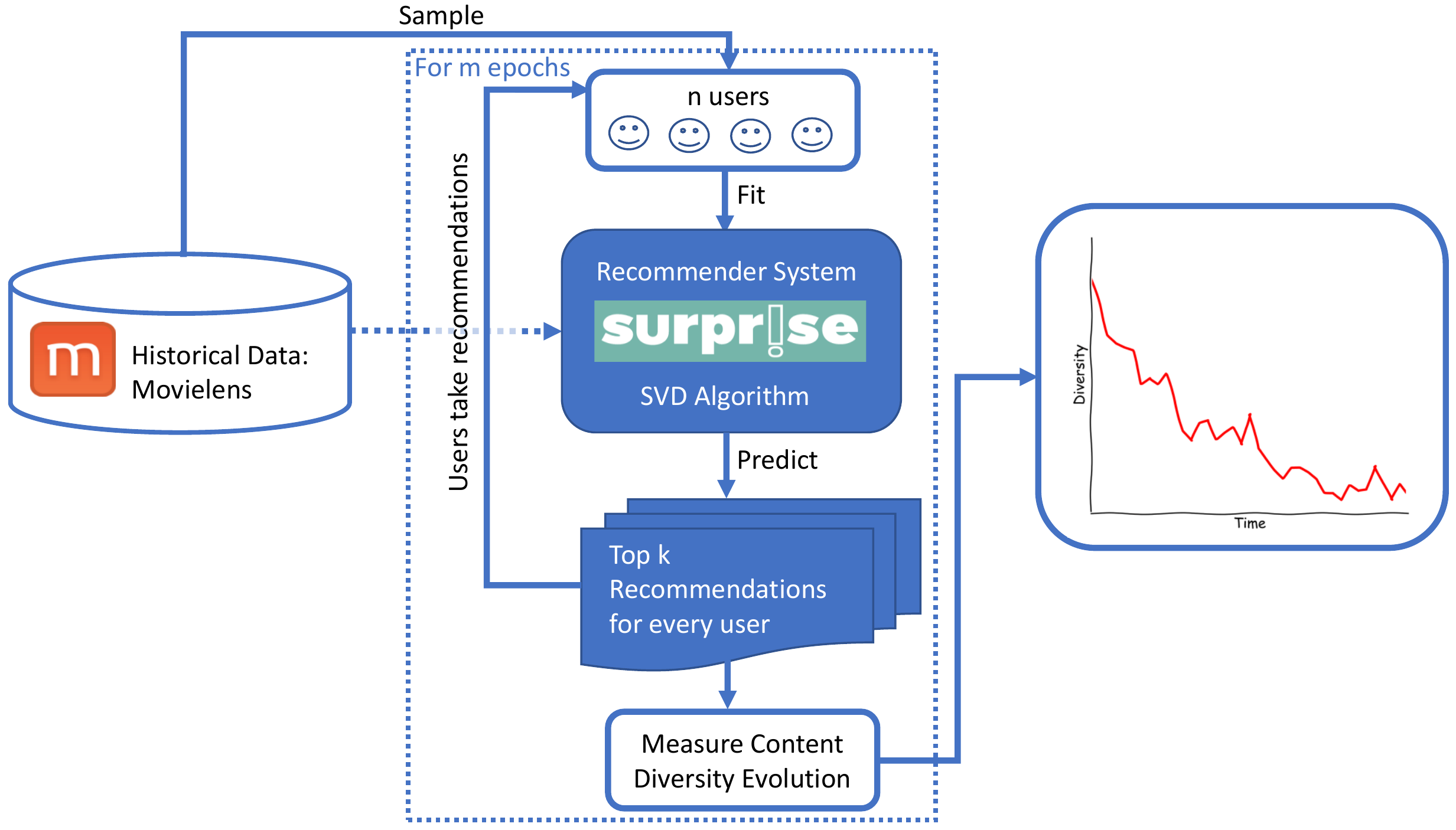}
\caption{First, the initial state of the system is fitted to historical data. Then, a population of users is sampled from the dataset to study the effect of their prolonged exposure to the system. The user-recommender interactions are modelled by recommending new movies to the users, appending the recommendations to the dataset and refitting the model, over a number of epochs. The assumption is made that the users take the recommendations exactly and discover $k$ movies every epoch. Last, the evolution of the content diversity is measured and analyzed.
}
\label{fig:flowchart}
\end{figure}

The experimental setup is presented in Figure \ref{fig:flowchart}. First, we create the initial state of the system by fitting the recommender model to historical data from the MovieLens dataset. Then, we sample a population of users from the dataset to study the effect of their prolonged exposure to the system. We model the users' interaction with the recommendation engine by iteratively recommending new movies to the users, appending the recommendations to the dataset and refitting the model on the appended dataset. It is important to note that we assume that the users take recommendations and give exactly the ratings the system predicts. We run the experiment for a large number of epochs to simulate the evolution in time and measure the long-term effects on the content diversity. 

In our simulations, we used the implementation of the SVD model from \texttt{surprise} package \cite{surprize} in \texttt{Python}. We experimented with multiple setups. We varied the number $n$ of simulated users between 10 and 1000 and observed an intuitive pattern that increasing the number of simulated users reduces the amount of noise in our diversity measurements and allows to see the diversity evolution more clearly. We also experimented with the number $k$ of discovered (recommended) movies per epoch (we tried $k$ between 12 and 100 movies) and found that more discovered movies per epoch increase the rate at which the system descends into echo chambers. Additionally, we varied the number of epochs between 10 and 400 to see the trends over different time ranges. We used a subsample of 1000 users from the MovieLens dataset as the initial historical data, this subsample had 94157 ratings in total. We empirically found that the recommender we used was robust to hyperparameter changes and therefore we used the default hyperparameters of the SVD model from \texttt{surprise}.

\subsection{Measuring content diversity}\label{sec:diversity}

We measure the similarity between two movies following \cite{nguyen2014exploring}. Each movie $m_i$ is represented in movie-tag space as a vector of length $N$, and there are $M$ movies in total. The relevance of a particular movie-tag $t_k$ to a given movie $m_i$ is represented as $\mathrm{rel}(t_k,m_i)$, and we denote the movie-tag relevance matrix as $R \in \mathcal{R}^{N \times M}$, where $R_{k,i}$ = rel($t_k, m_i$).

We compute the Euclidean distance (L2 distance) $d$ between two movies $m_i$ and $m_j$ in movie-tag space as:
\begin{equation}\label{eq:dist}
    d(m_i,m_j) = \sqrt{\sum_{k=1}^N \left( \mathrm{rel}(t_k,m_i)- \mathrm{rel}(t_k,m_j) \right)^2}
\end{equation}

Smaller distances represent movies that are more similar to one another based on their content, as expressed by their relevant movie-tags. The two most similar movies in the MovieLens dataset are ``Saw V (2008)" and ``Saw VI (2009)".


For a set of movies $\mathbf{m}$ we define the diversity $D(\mathbf{m})$ as the average Euclidean distance of all unique pairwise combinations of those movies. That is:

\begin{equation}\label{eq:diversity}
    D(\mathbf{m}) = \frac{2!( |\mathbf{m}| - 2)!}{|\mathbf{m}|!}\sum_{i=1}^{|\mathbf{m}|} \sum_{j>i}^{|\mathbf{m}|} d(m_i,m_j)
\end{equation}
where $|\mathbf{m}|$ is the number of movies in $\mathbf{m}$.



\subsection{Visualizing diversity}
\begin{figure*}[!htb]
\centering
\includegraphics[width=0.95\textwidth]{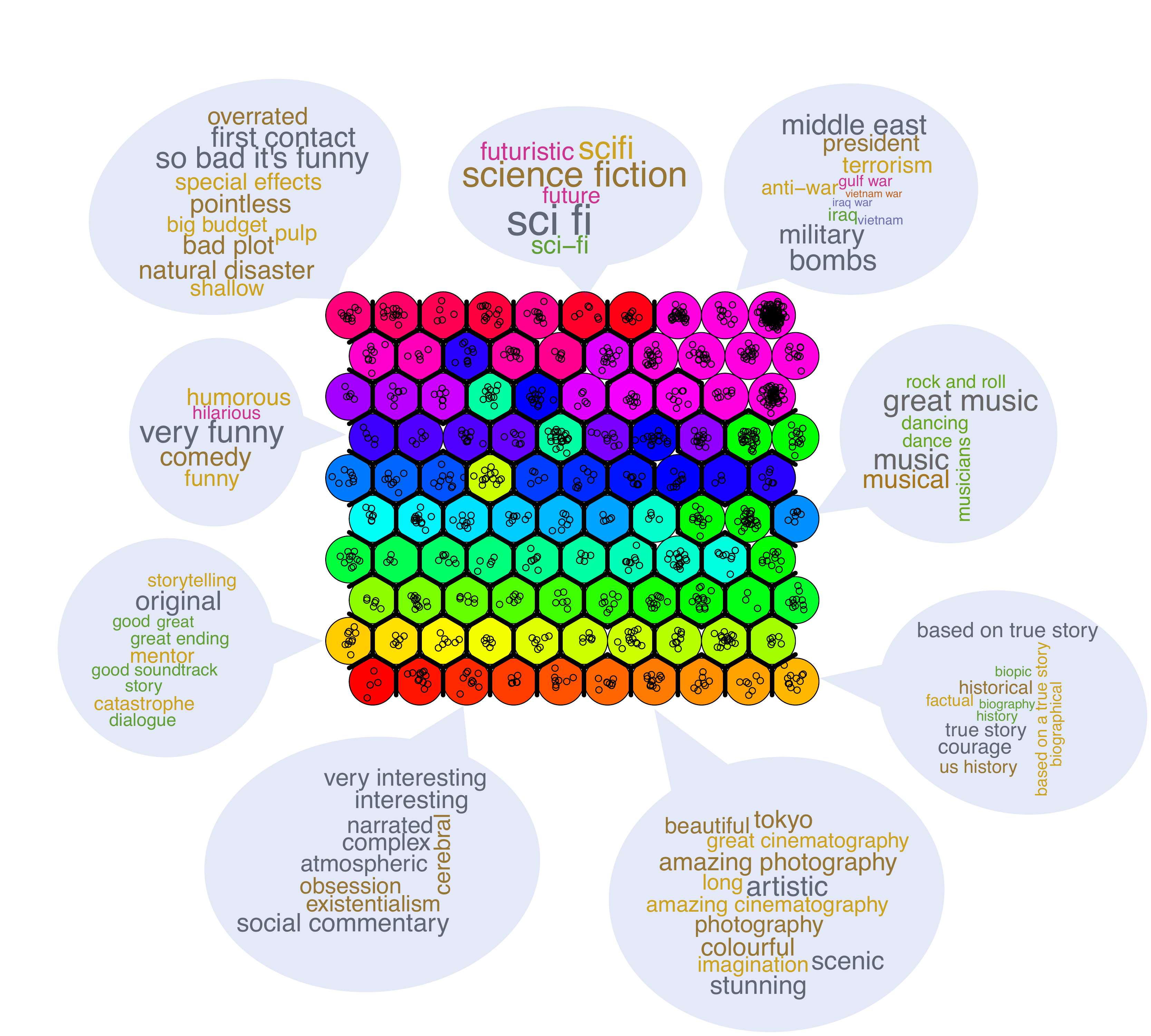}
\caption{Movie tags clustered by applying Self-Organising Map (SOM) on the movie-tag scores data. Two levels of clustering have been performed. 1) The movie-tags are clustered into a total of $100$ nodes where each node contains a set of movie-tags that are similar in content. The hollowed circles in a node refer to the movie-tags that are clustered together. Each word cloud around the SOM represents the tags clustered into one node. 2) The nodes are further grouped using hierarchical clustering, with the resulting groups highlighted in color. 
}
\label{fig:SOM_cloud}
\end{figure*}

\indent 
While the mean Euclidean distance in the movie-tag space is a good quantitative measure of diversity, it does not allow us to visualize how diversity changes over time. Self-Organizing Map (SOM) \cite{kohonen1982self}, work extremely well in this context. SOM is an unsupervised nonlinear dimensionality reduction and clustering algorithm based on artificial neural networks. SOM clusters tags and creates a 2-D representation of the tag clusters that can be used to visualize the diversity of any set of movies, and the recommended movies in particular (See Figure \ref{fig:SOM_cloud}).

Intuitively, if two tags are similarly relevant to all the movies, then they are more likely to be clustered into one node of SOM. The nine word clouds around the SOM show the content of sample tag clusters (nodes), with size of the tags in the word cloud indicating the average relevance for all the movies. Furthermore, nodes that are closer in SOM are more likely to be similar.


\indent
However, it is still possible that similar nodes appear in different areas on the map. To mitigate this, we further performed hierarchical clustering to group the nodes, and color-coded the resulting node groups on the SOM. 

Using SOM, we can visualize the diversity of a given set of movies by highlighting relevant tag nodes (see Figure \ref{fig:results}B,C). If the average relevance score of a tag for the given set of movies exceeds $80$th percentile of the relevance scores for that tag among all the movies, then the node containing that tag is highlighted. If a node contains no such tags, then it has no color (white).

\section{Results and Discussion}

\subsection{Shrinking diversity in content recommendation}

\begin{figure*}[!htb]
\centering
\includegraphics[width=0.95\textwidth]{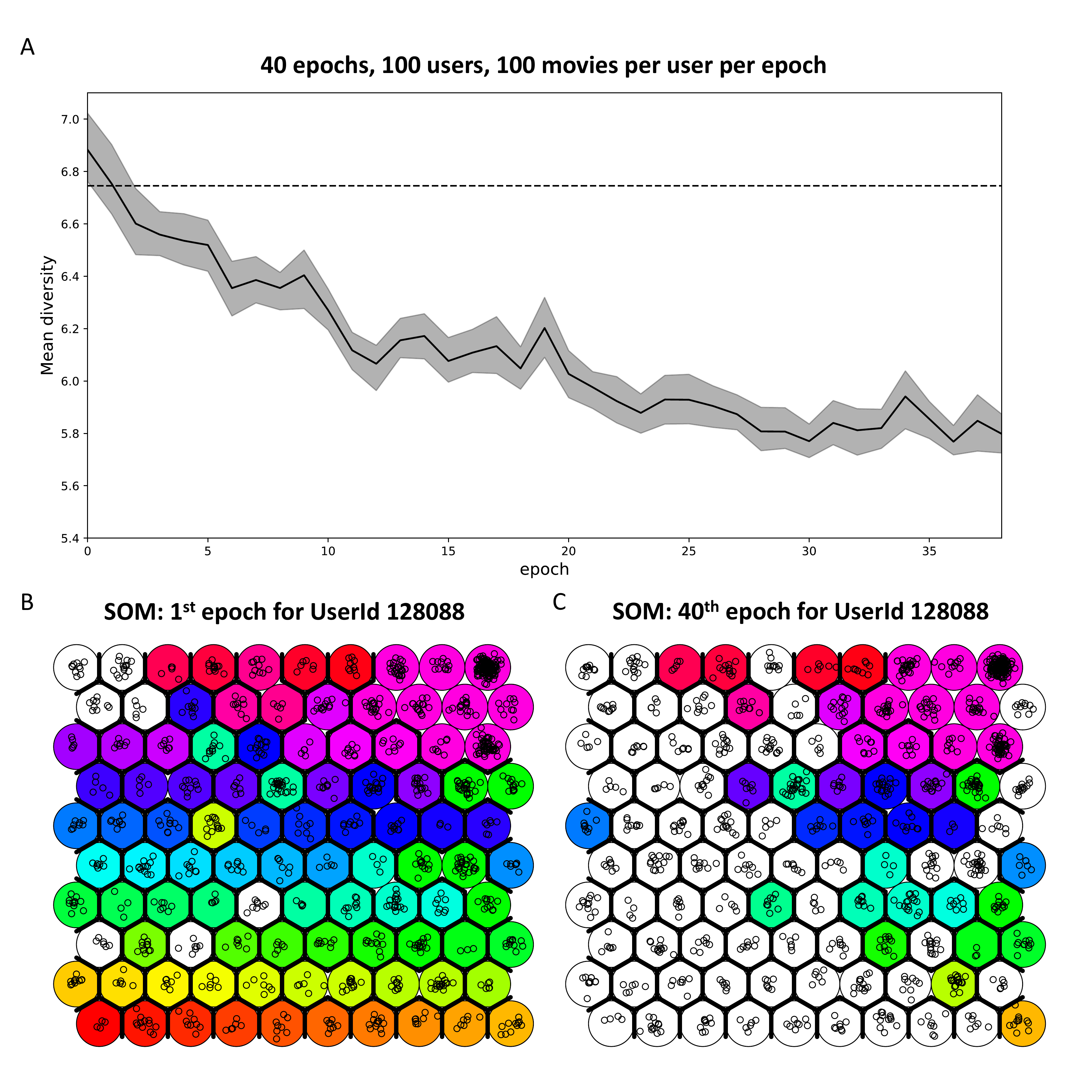}
\caption{A: The mean user diversity of the sample $\left<D\left(\mathbf{m}\right)\right>$ as a function of epoch is shown in solid black. The gray band shows the $1\sigma$ confidence interval of the user diversity at each epoch. The mean user diversity in the sample prior to initializing the recommendation system is shown as a horizontal dashed black line. There is a clear decrease in the diversity of content recommended by the model over time, a $15\sigma$ effect by epoch 40. B: The set of movies $\mathbf{m}$ that are recommended by the model in the first epoch are shown in the SOM latent space. Colored cells show clusters of movie-tags that are relevant to the set of movies $\mathbf{m}$ that are recommended by the model in the first epoch. C: Same as B but at the 40th epoch. There is again a clear decrease in the diversity of content recommended by the model, complementing the result in A. 
}
\label{fig:results}
\end{figure*}

In order to assess the impact of our recommendation system on content diversity, we iteratively simulate users interacting with the system as detailed in \cref{sec:sim}.
At each iteration (epoch) of this procedure, we compute the average diversity of the content recommended by the model (\cref{eq:diversity}) and visualize the associated movie-tags in the SOM latent space. This is shown in \cref{fig:results}. Here we see a distinct decline in the mean diversity of system recommendations over time. The mean user diversity prior to entering the recommendation system is $\left<D\left(\mathbf{m}\right)\right>=6.75\pm0.06$. By epoch 10 this value has dropped to $\left<D\left(\mathbf{m}\right)\right>=6.30\pm0.01$, a $7\sigma$ decrease in the mean diversity of content the users are exposed to.
Clearly the recommendation system is having a profound, statistically significant impact on the diversity of content suggested to the users.

The lower panels of \cref{fig:results} showcase the set of movies recommended by the system in the latent space of the SOM. This is shown for one random user in epoch 1 and in epoch 40. This visualization complements the result in panel A. When comparing to the movie-tag representation of the SOM cells given in \cref{fig:SOM_cloud}, we see that while this user was initially exposed to a wide variety of content by the recommendation system, by epoch 40 these recommendations are highly focused on content characterized by tags such as ``sci-fi" and ``war". 

This specific simulation utilized 100 movie recommendations per epoch over the course of 40 epochs. This is representative of the long term impact of a system where recommendations are given at a low cadence. We test the sensitivity of our result to variations in the number of epochs, movies per epoch, and users sampled and find that in all cases the observed contraction in content diversity is statistically significant.


These results illustrate the shrinking diversity of content that is recommended to users and the formation of ``echo-chambers" over long term exposure to recommendation systems. 
Perhaps the most damaging aspect of this effect is that it is invisible. The users of these systems are often unaware that they are being shown an ever-shrinking bubble of content that influences their perspectives and behaviours.

We believe this work can be directly used in practice to implement a simple application that allows users to monitor the diversity of content they are exposed to over time and stay cognisant of the extent of their echo chamber. 


\subsection{Repairing recommendation diversity}\label{sec:repair}

We tried to investigate whether or not an individual user can break themselves out of an echo chamber solely by manipulating their own content ratings. We framed this question as an optimization problem with the objective function given by the recommended content diversity of a user and the decision variables being the user's content ratings. 

However, the diversity of the recommended content can only be evaluated as a black-box function of user ratings (the algorithm for its evaluation is presented in Figure \ref{fig:flowchart}) and the underlying black-box optimization problem of maximizing the diversity turned out to be difficult and possibly infeasible to solve. 

We attempted to solve this optimization problem using three different approaches: 
\begin{enumerate}
    \item Optimization methods that attempt to approximate the black-box function derivative information using finite differences (e.g. constrained trust region methods \cite{conn2000trust} implemented in \texttt{scipy})
    \item Black-box optimization methods based on sampling techniques which attempt to minimize a black-box function in a derivative-free fashion (e.g. \cite{knysh2016blackbox}). 
    \item Heuristic approaches. We devised an intuitive heuristic solution based on our content diversity metric: adding high ratings for the movies that are as far as possible from the user's viewing history. Second, we experimented with a simple idea of adding high ratings to random movies to leverage the power of randomization.
\end{enumerate}
We observed that none of these approaches managed to significantly increase the objective function.

\section{Conclusion}

Recommendation systems are designed to enhance the user experience by predicting and suggesting content that they will enjoy. An unintended consequence of these systems is that they may decrease the diversity of content users are exposed to. In this paper, we quantitatively investigated the extent of this effect in collaborative filtering systems.
We did this by simulating the interaction of users in the MovieLens dataset with a recommendation system and monitoring changes in content diversity over time. Our primary findings are as follows:

\begin{enumerate}
    \item The diversity of content recommended by a collaborative filtering based model decreases significantly over time. In one simulation, we find that the diversity of the content recommended by the model decreases by $7\sigma$ within 10 iterations. Qualitatively our results are independent of the hyper-parameters governing our model and simulation.
    \item It is infeasible for a single user to break themselves out of the echo-chamber induced by the recommendation system when they can only manipulate their own ratings. 
\end{enumerate}






\bibliography{bibtex}

\begin{thebibliography}{10}

\bibitem{bobadilla2013recommender}
J.~Bobadilla, F.~Ortega, A.~Hernando, and A.~Guti{\'e}rrez, ``Recommender
  systems survey,'' {\em Knowledge-based systems}, vol.~46, pp.~109--132, 2013.

\bibitem{dandekar2013biased}
P.~Dandekar, A.~Goel, and D.~T. Lee, ``Biased assimilation, homophily, and the
  dynamics of polarization,'' {\em Proceedings of the National Academy of
  Sciences}, vol.~110, no.~15, pp.~5791--5796, 2013.

\bibitem{jiang2019degenerate}
R.~Jiang, S.~Chiappa, T.~Lattimore, A.~Gy{\"o}rgy, and P.~Kohli, ``Degenerate
  feedback loops in recommender systems,'' in {\em Proceedings of the 2019
  AAAI/ACM Conference on AI, Ethics, and Society}, pp.~383--390, 2019.

\bibitem{mcnee2006being}
S.~M. McNee, J.~Riedl, and J.~A. Konstan, ``Being accurate is not enough: how
  accuracy metrics have hurt recommender systems,'' in {\em CHI'06 extended
  abstracts on Human factors in computing systems}, pp.~1097--1101, 2006.

\bibitem{harper2015movielens}
F.~M. Harper and J.~A. Konstan, ``The movielens datasets: History and
  context,'' {\em Acm transactions on interactive intelligent systems (tiis)},
  vol.~5, no.~4, pp.~1--19, 2015.

\bibitem{vig2012tag}
J.~Vig, S.~Sen, and J.~Riedl, ``The tag genome: Encoding community knowledge to
  support novel interaction,'' {\em ACM Transactions on Interactive Intelligent
  Systems (TiiS)}, vol.~2, no.~3, pp.~1--44, 2012.

\bibitem{koren2009matrix}
Y.~Koren, R.~Bell, and C.~Volinsky, ``Matrix factorization techniques for
  recommender systems,'' {\em Computer}, vol.~42, no.~8, pp.~30--37, 2009.

\bibitem{Ricci_book}
N.~Manouselis, H.~Drachsler, R.~Vuorikari, H.~Hummel, R.~Koper, F.~Ricci,
  L.~Rokach, B.~Shapira, and P.~Kantor, ``Recommender systems handbook,'' {\em
  Recommender Systems in Technology Enhanced Learning}, pp.~387--415, 2011.

\bibitem{surprize}
N.~Hug, ``Surprise: A python library for recommender systems,'' {\em Journal of
  Open Source Software}, vol.~5, no.~52, p.~2174, 2020.

\bibitem{nguyen2014exploring}
T.~T. Nguyen, P.-M. Hui, F.~M. Harper, L.~Terveen, and J.~A. Konstan,
  ``Exploring the filter bubble: the effect of using recommender systems on
  content diversity,'' in {\em Proceedings of the 23rd international conference
  on World wide web}, pp.~677--686, 2014.

\bibitem{kohonen1982self}
T.~Kohonen, ``Self-organized formation of topologically correct feature maps,''
  {\em Biological cybernetics}, vol.~43, no.~1, pp.~59--69, 1982.

\bibitem{conn2000trust}
A.~R. Conn, N.~I. Gould, and P.~L. Toint, {\em Trust region methods}.
\newblock SIAM, 2000.

\bibitem{knysh2016blackbox}
P.~Knysh and Y.~Korkolis, ``Blackbox: A procedure for parallel optimization of
  expensive black-box functions,'' {\em arXiv preprint arXiv:1605.00998}, 2016.

\end{thebibliography}






\end{document}